\documentclass[aps,prd,floatfix,superscriptaddress,nofootinbib,longbibliography,reprint]{revtex4-2}
\usepackage{amsmath,amssymb,bm,graphicx,booktabs,multirow,siunitx,microtype,xcolor,hyperref}
\hypersetup{colorlinks=true,linkcolor=blue,citecolor=blue,urlcolor=blue}
\usepackage{url}
\usepackage{xspace}
\DeclareRobustCommand{\NNLOJET}{NNLO\scalebox{.8}{JET}\xspace}
\usepackage{fontawesome5,scalerel,mathrsfs,lineno,cancel}
\usepackage{scalerel}
\usepackage{mathrsfs}
\usepackage{cancel}
\usepackage{tikz}
\usepackage[normalem]{ulem}
\makeatletter

\newcommand{\Rmnum}[1]{\expandafter\@slowromancap\romannumeral #1@}
\makeatother
\begin{document}

\title{Entanglement of back-to-back gluon pair}

\author{Lei Yang}
\affiliation{Institute of Frontier and Interdisciplinary Science, Key Laboratory of Particle Physics and Particle Irradiation (MOE), Shandong University, Qingdao, Shandong 266237, China}

\author{Wen-Hao Yao}
\affiliation{Institute of Frontier and Interdisciplinary Science, Key Laboratory of Particle Physics and Particle Irradiation (MOE), Shandong University, Qingdao, Shandong 266237, China}

\author{Yu-Kun Song}
\email{sps\_songyk@ujn.edu.cn}
\affiliation{School of Physics and Technology, University of Jinan, Jinan, Shandong 250022, China}

\author{Shu-Yi Wei}
\email{shuyi@sdu.edu.cn}
\affiliation{Institute of Frontier and Interdisciplinary Science, Key Laboratory of Particle Physics and Particle Irradiation (MOE), Shandong University, Qingdao, Shandong 266237, China}

\begin{abstract}
We investigate the polarization correlation of back-to-back gluon pairs in unpolarized $pp$ collisions and compute their density matrix. For gluon pairs produced through $q\bar{q}$ annihilation, the linear polarization correlation is sizable, even reaching unity when the final-state gluons are emitted perpendicular to the beam direction in the $q\bar{q}$ rest frame. Combining with the fact that their helicity remains maximally correlated across different scattering angle, the gluon pair system thus resides in a maximally entangled Bell state in this configuration. In contrast, for the $gg\to gg$ channel, both the helicity and linear polarization correlations are considerably suppressed, producing a separable state. Nonetheless, it still retains an appreciable linear polarization correlation of about $11\%$ at central scattering in unpolarized collisions. The linearly polarization correlation of the back-to-back gluon pair can further increase when the incoming gluons are circularly polarized. Employing the anisotropy of energy correlators, we demonstrate that the linear polarization correlation of gluon pairs can be extracted from the $\cos 2\Delta\Phi$ modulation.
\end{abstract}

\maketitle

\section{Introduction}

Quantum entanglement is a defining feature of quantum mechanics~\cite{Bell:1964kc,Clauser:1969ny,Cirelson:1980ry,Peres:1996dw,Horodecki:1997vt,Horodecki:2009zz}. In high energy collisions, the final-state partons can acquire nontrivial spin correlation even if the colliding particles are unpolarized. The spin correlation can approach unity in certain phase space, and therefore, the final-state bipartite system becomes maximally entangled~\cite{Garrido:2025xpk,Liu:2026dzv}. This feature not only allows us to investigate the Bell nonlocality in high energy collisions \cite{Gong:2021bcp,Fabbrichesi:2023idl,Wu:2024asu,Guo:2024jch,Barr:2024djo,Fabbrichesi:2024rec,Fabbrichesi:2025aqp,Afik:2025ejh,Qi:2025onf,Cheng:2025cuv,Cao:2025qua,Zhang:2025ean,Zhang:2026rnl,Bloss:2026yrf,Fabbrichesi:2026jvm,Zhang:2026pff,Hatta:2026dqs}, but also opens a new avenue for a variety of intriguing topics, such as the decoherence effect~\cite{Gu:2025ijz,Lin:2025eci}, the hadronization of polarized partons \cite{Li:2023qgj,Zhang:2023ugf,Datta:2024hpn,Chen:2024qvx,Yang:2024kjn,Huang:2024awn,Silva:2025dan,Xi:2025feb} and global polarization of QGP~\cite{Lv:2024uev,Shen:2024buh,Chen:2024hki,Zhang:2024hyq,Sheng:2025puj,Oliva:2026wbo,Lv:2026kct}. 

Over the last decades, extensive studies have explored the spin correlation within spin-$1/2$ bipartite systems, such as the $\Lambda\bar{\Lambda}$ or $t\bar{t}$ pairs \cite{Tornqvist:1980af,Chen:1994ar,Bernreuther:2001rq,Bernreuther:2004jv,Bernreuther:2006vg,Bernreuther:2010ny,Bernreuther:2013aga,Bernreuther:2015yna,Bernreuther:2017yhg,Afik:2020onf,Fabbrichesi:2021npl,Severi:2021cnj,Afik:2022kwm,Ehataht:2023zzt,Fabbrichesi:2024xtq,Bernreuther:2024ltu,Fabbrichesi:2024wcd,Fabbrichesi:2025psr,Liu:2026ees,Barata:2026hck,Zhang:2026tqs}. Their weak decays provide an experimentally accessible spin analyzer, allowing the spin correlation to be reconstructed from the angular distribution of their decay products. Recent measurements in different experimental facilities \cite{ATLAS:2012ao,BESIII:2018cnd,CMS:2019nrx,ATLAS:2019zrq,ATLAS:2023fsd,CMS:2024pts,CMS:2024zkc,BESIII:2025vsr,STAR:2025njp,CMS:2026edg} demonstrate that these studies are not only of theoretical interest per se, but also well within current experimental reach. As a joint effort, the study of spin correlations/quantum entanglement has emerged as a novel frontier recently.

While extensive literature has focused on fermions, recent efforts have increasingly turned to the spin correlations of gauge bosons~\cite{Barr:2021zcp,Aguilar-Saavedra:2022wam,Ashby-Pickering:2022umy,Fabbrichesi:2022ovb,Fabbrichesi:2023cev,Fabbrichesi:2023jep,Goncalves:2025mvl,Goncalves:2025xer,Goncalves:2026njf,Pan:2026xgx,Pan:2026usx,Cheng:2026ktp,deLima:2026jbb}, such as $W$, $Z$, and photons, which can also be measured in experiments~\cite{CMS:2021icx,ATLAS:2022oge,ATLAS:2026hye}. These studies offer fresh perspectives on quantum entanglement at high energy colliders. In contrast, the gluon sector receives relatively less attention. Investigating this regime is equally important, since it offers deep insight into the spin dynamics of the strong interaction. To bridge this gap, we compute the density matrix of gluon pairs produced in unpolarized hadron colliders and analyze their entanglement properties. 

The energy correlator~\cite{Basham:1978bw,Basham:1978zq} has emerged as a powerful probe of the angular structure of energy flow in high energy processes, with applications spanning a broad range of QCD physics \cite{Belitsky:2013ofa,Moult:2018jzp,Dixon:2018qgp,Kologlu:2019mfz,Gao:2019ojf,Korchemsky:2019nzm,Dixon:2019uzg,Luo:2019nig,Chen:2020vvp,Chen:2020adz,Ebert:2020sfi,Chen:2021gdk,Duhr:2022yyp,Komiske:2022enw,Neill:2022lqx,Chen:2022jhb,Li:2021txc,Liu:2022wop,Kang:2023big,Chen:2024bpj,Alvarez:2023fhi,Li:2023gkh,Boussarie:2023izj,Chen:2023zzh,Chen:2023wah,Kang:2023gvg,Liu:2023aqb,Chen:2024nyc,Guo:2024vpe,Liu:2024lxy,Moult:2025nhu,Chen:2025rjc,Chang:2025kgq,Monni:2025zyv,Gao:2025evv,Cao:2025icu,Jaarsma:2025tck,Song:2025bdj,Gonzalez:2026dzy,Wang:2026uej,Ruan:2026xyd}. Their sensitivity to spin dependent angular structures also provides access to polarization information of the initial parton~\cite{Chen:2020adz,Li:2021txc,Liu:2022wop,Li:2023gkh,Kang:2023big,Chen:2024bpj,Song:2025bdj}. The gluon linear polarization has also been extensively studied in complementary contexts. The idea of probing gluon linear polarization from the azimuthal angle anisotropy can be traced back to Ref.~\cite{Bacchetta:2004it}. Within the TMD and CGC frameworks, linearly polarized gluon distributions in unpolarized hadrons and nuclei can generate characteristic azimuthal modulations~\cite{Bacchetta:2004it, Boer:2010zf,Metz:2011wb,Dominguez:2011br,Akcakaya:2012si,Boer:2017xpy,He:2026aeh}. Therefore, the linear polarization correlation of gluon pairs can also be measured through the anisotropy of energy correlators. In particular, a previous work showed that the linear polarization of a single gluon can be extracted from the $\cos 2\phi$ azimuthal anisotropy of energy correlations within a gluon jet~\cite{Song:2025bdj}. 

In this work, we investigate the polarization correlations of gluon pairs produced in the $q\bar{q}\to gg$ and $gg\to gg$ channels by constructing their complete polarization density matrices and analyzing the corresponding entanglement properties. The spin dependent collinear splittings of the gluon pair map this spin correlation onto a characteristic $\cos 2\Delta\Phi$ modulation~\cite{Collins:1987cp,Ellis:1996mzs,Karlberg:2021kwr,Song:2025bdj,Wang:2026uej}, where $\Delta\Phi$ denotes the difference between the azimuthal angles of the two branching planes, each measured with respect to the scattering plane. Established on this property, we employ the four-point energy correlator (E4C) to probe the linear polarization correlation of the gluon pair. 

The rest of this paper is organized as follows. In Sec.~\ref{sec:formalism}, we introduce the density matrix formalism for the gluon pair and its Bloch decomposition. In Sec.~\ref{sec:partonic-density}, we evaluate the density matrix of the gluon pair produced in different channels and analyze their polarization correlations as well as entanglement properties. In Sec.~\ref{sec:phenomenology}, we show how the four-point energy correlator probes the linear polarization correlation through the azimuthal anisotropy. A brief summary is given in Sec.~\ref{sec:summary}.

\section{Density Matrix Formalism}\label{sec:formalism}

We work in the center-of-mass frame of the final-state gluon pair. At the leading order approximation, this is also the rest frame of the colliding partons. The initial and final-state partons thus form the scattering plane. For the first gluon, its momentum direction $\hat{\bm{k}} = \bm{e}_z$ defines the local longitudinal axis and the associated transverse plane. We therefore choose an orthogonal transverse frame, with the transverse plane spanned by $\bm{e}_x$ and $\bm{e}_y$
\begin{align}
    \bm{e}_x\cdot\hat{\bm{k}}=\bm{e}_y\cdot\hat{\bm{k}}=0,
    \quad\bm{e}_x\cdot\bm{e}_y=0, 
    \quad\bm{e}_x\times\bm{e}_y=\hat{\bm{k}},
    \label{eq:local-transverse-basis}
\end{align}
where $\bm e_x$ and $\bm{e}_y$ are transverse unit vectors perpendicular to $\hat{\bm{k}}$. While $\bm{e}_x$ lies in the scattering plane, $\bm{e}_y$ is normal to it. For the other gluon, we define the local axis in the same way and obtain
\begin{align}
    &  \bm{e}_{z'} = - \bm{e}_z,
    && \bm{e}_{x'} = - \bm{e}_x,
    && \bm{e}_{y'} = + \bm{e}_y.
    \label{eq:local-basis-relation}
\end{align}
A linearly polarized gluon with polarization angle $\phi$, measured from its local axis $\bm{e}_x$ in the local transverse plane, has the polarization vector
\begin{align}
    \boldsymbol \varepsilon(\phi) = \cos\phi\,\bm{e}_x+\sin\phi\,\boldsymbol e_y.
    \label{eq:linear-polarization-vector}
\end{align}

For a gluon with fixed momentum $k$, gauge invariance leaves only two physical helicity states, $|k,\lambda\rangle$ with $\lambda=\pm1$. Ward identities ensure that gauge dependent components of the polarization vector do not contribute to physical observables. The physical polarization state is therefore described by a $2\times 2$ density matrix $\rho_{\lambda \lambda'}$ in the helicity basis, which can be further decomposed in terms of Pauli matrices $\sigma_i$ with $i=x,y,z$ and unit matrix $\mathbb{I}$. $\sigma_z$ measures the circular polarization of the gluon, whereas $\sigma_x$ and $\sigma_y$ characterize the two independent components of linear polarization. Although we have used the same decomposition form as that for spin-$1/2$ particles, the physical interpretations of $\sigma_{x,y}$ discussed below are quite different, reflecting the fact that we are merely using the Pauli matrices as a tool for simplicity. Equivalently, the polarization state can be written in covariant form as~\cite{Berestetskii:1982qgu,Barr:2024djo}
\begin{align}
    \widetilde\rho_{\mu\nu}
    =
    \sum_{\lambda,\lambda'=\pm}
    \rho_{\lambda\lambda'}\,
    \varepsilon_\mu(k,\lambda)\varepsilon_\nu^*(k,\lambda').
    \label{eq:covariant-polarization-density}
\end{align}

For a two gluon final state, the physical helicity spaces of the two gluons form a bipartite qubit system. The corresponding density matrix admits the following Bloch decomposition~\cite{Bloch:1946zza,Fano:1957zz,Fano:1983zz}: 
\begin{align} 
    \rho = \frac{1}{4} \Biggl[ & \mathbb{I} \otimes \mathbb{I} 
    +\sum_{i=x,y,z} B_i^+ (\sigma_i\otimes \mathbb{I}) +\sum_{j'=x',y',z'} B_{j'}^-(\mathbb{I} \otimes\sigma_{j'}) 
    \nonumber\\
    & 
    +\sum_{i, j'}^3 C_{ij'}(\sigma_i\otimes \sigma_{j'}) \Biggr], 
    \label{eq:two-gluon-decomposition} 
\end{align} 
where $B_i^+=\mathrm{Tr}[\rho(\sigma_i\otimes \mathbb{I})]$, $B_{j'}^-=\mathrm{Tr}[\rho(\mathbb{I} \otimes \sigma_{j'})]$, and $C_{ij'}=\mathrm{Tr}[\rho(\sigma_i\otimes \sigma_{j'})]$. The coefficients $B_i^\pm$ characterize the polarization of the gluon pair in their respective local frames. Specifically, $B_x^\pm$ quantifies the linear polarization asymmetry between $\phi=0$ and $\phi=\pi/2$, while $B_y^\pm$ measures that between $\phi=\pi/4$ and $\phi=3\pi/4$. The coefficients $C_{ij'}$ form the $3\times3$ polarization correlation tensor and encode correlations between polarization sensitive observables of the two gluons. 

To give a probabilistic interpretation of these coefficients, we define the joint projection probability
\begin{align}
    P(\phi_1,\phi_2^\prime)
    = \mathrm{Tr}\left[\rho\,\Pi(\phi_1)\otimes\Pi(\phi_2^\prime)\right],
    \label{eq:joint-stokes-response}
\end{align}
where $\Pi(\phi)=|\phi\rangle\langle\phi|$ projects onto a linear polarization state at angle $\phi$. The angles $\phi_1$ and $\phi_2^\prime$ are defined in the local transverse planes orthogonal to $\hat{\bm k}_1$ and $\hat{\bm k}_2$, respectively. Joint helicity probabilities are defined analogously using $\Pi(\lambda)=|\lambda\rangle\langle\lambda|$, with $\lambda=\pm1$ denoting helicity along the corresponding gluon momentum. A related discussion for photon systems also appeared recently~\cite{deLima:2026jbb}.

\begin{figure}[htb]
    \centering
    \includegraphics[width=0.9\linewidth]{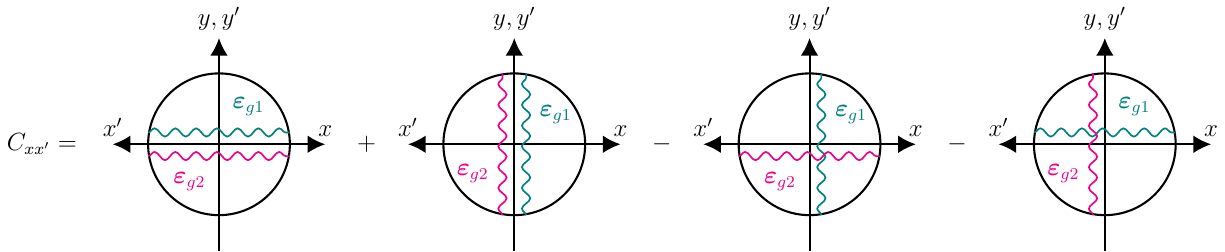}
    \vspace{0.4cm}
    \includegraphics[width=0.9\linewidth]{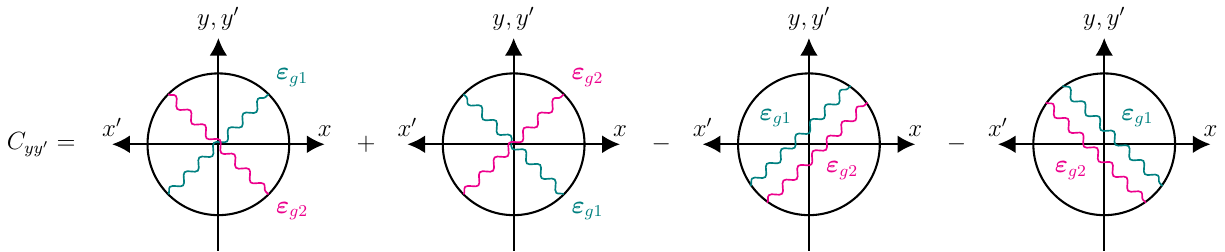}
    \caption{Illustrations of the linear polarization correlations $C_{xx'}$ and $C_{yy'}$. The teal and magenta lines represent the polarization axis of the final state gluons.}
    \label{fig:polarization-illustrations}
\end{figure}

For the diagonal elements relevant to the channels discussed below, one finds
\begin{align}
    &C_{xx'} = P(0,0) + P(\tfrac{\pi}{2},\tfrac{\pi}{2})
    - P(0,\tfrac{\pi}{2}) - P(\tfrac{\pi}{2},0),
    \label{eq:Cxx-prob}\\
    &C_{yy'} = P(\tfrac{\pi}{4},\tfrac{\pi}{4})
    + P(\tfrac{3\pi}{4},\tfrac{3\pi}{4})
    - P(\tfrac{\pi}{4},\tfrac{3\pi}{4})
    - P(\tfrac{3\pi}{4},\tfrac{\pi}{4}),
    \label{eq:Cyy-prob}\\
    &C_{zz'} = P(+,+) + P(-,-) - P(+,-) - P(-,+).
    \label{eq:Czz-prob}
\end{align}
Here $C_{xx'}$ measures the correlation between linear polarizations along the $\bm{e}_x$ and $\boldsymbol e_y$ axes, while $C_{yy'}$ measures the corresponding correlation in the basis rotated by $\pi/4$.\footnote{For instance, $C_{xx'} = 1$ implies that if the first gluon is polarized along the $x$-axis (or $y$-axis), the second gluon is also polarized along the $x'$-axis (or $y'$-axis), meaning their polarizations are parallel. Conversely, $C_{yy'} = 1$ indicates that if the first gluon is polarized along the bisector of $x$ and $y$ axes, the other is polarized along the bisector of $x'$ and $y'$ axes. Given that $\bm{e}_x = -\bm{e}_{x'}$ and $\bm{e}_y = \bm{e}_{y'}$, these two polarizations are perpendicular to each other.} The coefficient $C_{zz'}$ measures the excess of same-helicity over opposite-helicity configurations. Thus, $C_{xx'}$ and $C_{yy'}$ characterize linear polarization correlations and are sensitive to helicity coherence, whereas $C_{zz'}$ characterizes the helicity correlation. The physical interpretation of $C_{xx'}$ and $C_{yy'}$ has been illustrated in Fig.~\ref{fig:polarization-illustrations}.

\section{Density Matrix of the back-to-back gluon pair in unpolarized $pp$ collisions}\label{sec:partonic-density}

In this section, we investigate the polarization state of the two gluon system produced in hard scattering. By retaining the final-state helicity indices, we construct the corresponding polarization density matrices for the $q\bar{q}\to gg$ and $gg\to gg$ subprocesses at tree level, averaging over the helicities and colors of the initial-state partons and summing over the unobserved final-state colors. The resulting density matrices determine the polarization correlations generated in each channel and allow us to assess whether the corresponding two gluon state is entangled.

\subsection{Helicity density matrix}
\label{subsec:Helicity-density-matrix}

For a given partonic channel $ab\to gg$, the polarization state of the final-state gluon pair is described by the normalized helicity density matrix
\begin{align}
    \rho_{ab\to gg}=\frac{H_{ab\to gg}}{\operatorname{Tr}H_{ab\to gg}},
    \label{eq:rho-from-hard}
\end{align}
where $H_{ab\to gg}$ is the unnormalized helicity space hard matrix.  Its matrix elements are constructed from the partonic helicity amplitudes as
\begin{align}
    H^{\lambda_3\lambda_4,\lambda_3'\lambda_4'}_{ab\to gg}
    \equiv
    \sum_{\lambda_1,\lambda_2}
    \mathcal{M}_{\lambda_1\lambda_2;\lambda_3\lambda_4}
    \mathcal{M}^{*}_{\lambda_1\lambda_2;\lambda_3'\lambda_4'}.
    \label{eq:hard-matrix}
\end{align}
We represent the hard matrix in the basis $\{|++\rangle,|+-\rangle,|-+\rangle,|--\rangle\}$, where $|\pm\rangle$ are the helicity states defined in Eqs.~\eqref{eq:local-transverse-basis}--\eqref{eq:linear-polarization-vector} with phases fixed relative to the scattering plane. The partonic helicity amplitude calculations below follow the results of Ref.~\cite{Gastmans:1990xh}.

\subsubsection{\texorpdfstring{$q\bar{q}\to gg$}{q qbar to gg} channel}
\label{subsec:qqbar-channel}

For the quark annihilation channel, the normalized density matrix reads
\begin{align}
    \rho_{q\bar{q}\to gg}=
    \begin{pmatrix}
    0 & 0 & 0 & 0\\
    0 & \frac{1}{2} & \rho^{q\bar{q}}_{23} & 0\\
    0 & \rho^{q\bar{q}}_{32} & \frac{1}{2} & 0\\
    0 & 0 & 0 & 0
    \end{pmatrix}.
    \label{eq:rho-qqbar}
\end{align}
For massless quarks, chirality conservation fixes the helicity along the same fermion line. Together with the MHV/anti-MHV structure of the four-point $q\bar{q}gg$ amplitudes~\cite{Parke:1986gb,Mangano:1990by}, this gives $\mathcal{M}_{\lambda_1\lambda_2;++}=\mathcal{M}_{\lambda_1\lambda_2;--}=0$ for every allowed initial-state helicity configuration, while $\mathcal{M}_{\lambda_1\lambda_2;+-}$ and $\mathcal{M}_{\lambda_1\lambda_2;-+}$ can be nonzero. Hence, the rows and columns of $\rho_{q\bar{q}\to gg}$ associated with $|++\rangle$ and $|--\rangle$ vanish. 

The exchange symmetry of the two identical final-state gluons gives $H_{22}^{q\bar{q}}=H_{33}^{q\bar{q}}$. Since these are the only nonzero diagonal elements and $\operatorname{Tr}\rho_{q\bar{q}\to gg}=1$, one obtains $\rho_{22}^{q\bar{q}}=\rho_{33}^{q\bar{q}}=1/2$. The remaining off-diagonal elements are
\begin{align}
    \rho^{q\bar{q}}_{23}=\rho^{q\bar{q}}_{32}
    =\frac{H_{23}^{q\bar{q}}}{\operatorname{Tr}H_{q\bar{q}\to gg}}
    =-\frac{\sin^2\theta}{2(1+\cos^2\theta)},
    \label{eq:rho23-qqbar}
\end{align}
where $\theta$ denotes the scattering angle in the partonic center-of-mass frame.
These off-diagonal elements encode the coherence between $|+-\rangle$ and $|-+\rangle$. The corresponding nonzero $2\times2$ block is diagonalized by the symmetric and antisymmetric superpositions of these two states. We introduce the Bell basis
\begin{align}
    |\Phi^\pm\rangle=\frac{|++\rangle\pm|--\rangle}{\sqrt2},\quad
    |\Psi^\pm\rangle=\frac{|+-\rangle\pm|-+\rangle}{\sqrt2}.
    \label{eq:bell-basis}
\end{align}
Since the same-helicity configurations $|++\rangle$ and $|--\rangle$ are absent, the polarization state is entirely restricted to the $|\Psi^\pm\rangle$ sector of the Bell basis. In terms of these Bell states, the density matrix can be written as
\begin{align}
    \rho_{q\bar{q}\to gg}
    =\frac{1}{1+\cos^2\theta}|\Psi^-\rangle\langle\Psi^-|
    +\frac{\cos^2\theta}{1+\cos^2\theta}|\Psi^+\rangle\langle\Psi^+|.
    \label{eq:rho-qqbar-bell}
\end{align}

The Bell state weights make the angular evolution of the polarization state explicit. For central scattering, $\theta=\pi/2$, the weight of $|\Psi^-\rangle$ becomes unity, and the two gluon system reduces to the pure Bell state $|\Psi^-\rangle$. In the forward and backward limits, the two weights approach $1/2$, the coherence vanishes, and the state approaches an equal, separable mixture of the two opposite helicity configurations.

\subsubsection{\texorpdfstring{$gg\to gg$}{gg to gg} channel}
\label{subsec:gg-channel}

For gluon fusion, the normalized density matrix reads
\begin{align}
    \rho_{gg\to gg}=
    \begin{pmatrix}
    \rho_{11}^{gg} & 0 & 0 & 0\\
    0 & \rho_{22}^{gg} & \rho_{23}^{gg} & 0\\
    0 & \rho_{32}^{gg} & \rho_{33}^{gg} & 0\\
    0 & 0 & 0 & \rho_{44}^{gg}
    \end{pmatrix}.
    \label{eq:rho-gg}
\end{align}
Similarly, the MHV/anti-MHV structure of the four-gluon tree amplitudes leaves $H_{23}^{gg}=H_{32}^{gg}$ as the only nonzero off-diagonal elements, while all four diagonal helicity populations remain nonzero. Parity and gluon-exchange symmetries give $H_{11}^{gg}=H_{44}^{gg}$ and $H_{22}^{gg}=H_{33}^{gg}$. The normalized density matrix elements are
\begin{align}
    \rho_{11}^{gg}=\rho_{44}^{gg}
    &=\frac{H_{11}^{gg}}{2(H_{11}^{gg}+H_{22}^{gg})}
    =\frac{4}{(3+\cos^2\theta)^2},\\
    \rho_{22}^{gg}=\rho_{33}^{gg}
    &=\frac{H_{22}^{gg}}{2(H_{11}^{gg}+H_{22}^{gg})}
    =\frac{1+6\cos^2\theta+\cos^4\theta}
    {2(3+\cos^2\theta)^2},\\
    \rho_{23}^{gg}=\rho_{32}^{gg}
    &=\frac{H_{23}^{gg}}{2(H_{11}^{gg}+H_{22}^{gg})}
    =\frac{\sin^4\theta}{2(3+\cos^2\theta)^2}.
    \label{eq:rho-components-gg}
\end{align}

In the Bell basis of Eq.~\eqref{eq:bell-basis}, the density matrix reads
\begin{align}
    \rho_{gg\to gg}={}&\rho_{11}^{gg}
    \left(|\Phi^+\rangle\langle\Phi^+|
    +|\Phi^-\rangle\langle\Phi^-|\right)
    \nonumber\\
    &+(\rho_{22}^{gg}+\rho_{23}^{gg})
    |\Psi^+\rangle\langle\Psi^+| 
    \nonumber\\
    &+(\rho_{22}^{gg}-\rho_{23}^{gg})
    |\Psi^-\rangle\langle\Psi^-|.
    \label{eq:rho-gg-bell}
\end{align}    
Unlike the $q\bar{q}\to gg$ channel, the $gg\to gg$ final state never reduces to a pure Bell state because the two components $|\Phi^\pm\rangle$ have equal and nonzero weights, $p_{\Phi^+}=p_{\Phi^-}=\rho_{11}^{gg}>0$, for every scattering angle. For central scattering, $\theta=\pi/2$, the weights are $p_{\Phi^\pm}=4/9$, $p_{\Psi^+}=1/9$, and $p_{\Psi^-}=0$. In the forward and backward limits, $\theta\to0,\pi$, all four weights approach $1/4$. Thus, $\rho_{gg\to gg}$ remains mixed throughout the physical range of scattering angles and becomes maximally mixed in the forward and backward limits.

\subsection{Entanglement}
\label{subsec:Entanglement}

For both subprocesses, the individual outgoing gluons are unpolarized, and the nontrivial polarization information resides entirely in the two gluon correlations. The density matrices obtained above give the correlation tensors
\begin{align}
    C^{q\bar{q}}&=\operatorname{diag}(c_q,c_q,-1),\\
    C^{gg}&=\operatorname{diag}(c_g,c_g,c_z),
    \label{eq:correlation-tensors}
\end{align}
with
\begin{align}
    c_q(\theta)&=\rho_{23}^{q\bar{q}}+\rho_{32}^{q\bar{q}}
    =-\frac{\sin^2\theta}{1+\cos^2\theta},\\
    c_g(\theta)&=\rho_{23}^{gg}+\rho_{32}^{gg}
    =\frac{\sin^4\theta}{(3+\cos^2\theta)^2},\\
    c_z(\theta)&=\rho_{11}^{gg}-\rho_{22}^{gg}-\rho_{33}^{gg}+\rho_{44}^{gg}\nonumber\\
    &=\frac{\sin^2\theta(7+\cos^2\theta)}
    {(3+\cos^2\theta)^2}.
    \label{eq:correlation-coefficients}
\end{align}

\begin{figure}[htb]
    \centering
    \includegraphics[width=0.4\textwidth]{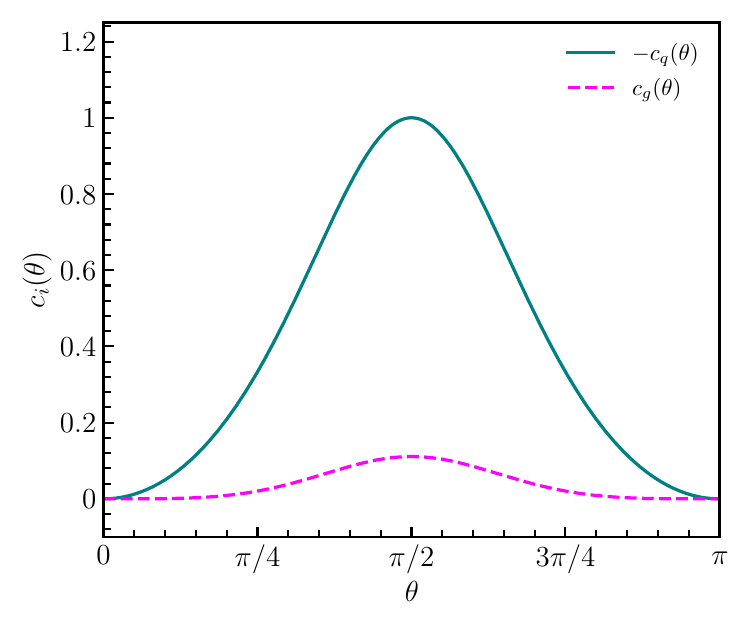}
    \caption{Angular dependence of the linear polarization correlations $-c_q(\theta)$ for $q\bar{q}\to gg$ and $c_g(\theta)$ for $gg\to gg$. Both peak at $\theta=\pi/2$, with maxima $1$ and $1/9$, respectively.}
    \label{fig:Cij_comparison}
\end{figure}

The numerical results for the angular dependence of linear polarization correlations in these two channels are shown in Fig.~\ref{fig:Cij_comparison}. The Bell basis decompositions derived above allow the entanglement content of these correlations to be quantified directly. Since both density matrices are diagonal in this basis, their entanglement is determined by the largest eigenvalue in the Bell basis. The concurrence~\cite{Wootters:1997id} is
\begin{align}
    \mathcal C=\max(0,2p_{\max}-1),
    \label{eq:concurrence-bell-diagonal}
\end{align}
where $p_{\max}$ denotes the largest of the four Bell-state probabilities. The state is therefore entangled if and only if $p_{\max}>1/2$.  The stronger condition of CHSH nonlocality is tested independently by the Horodecki criterion~\cite{Horodecki:1995nsk},
\begin{align}
    S_{\max}=2\sqrt{\mu_1+\mu_2},
    \label{eq:horodecki-criterion}
\end{align}
where $\mu_{1,2}$ are the two largest eigenvalues of $C^{\mathrm T}C$, and $S_{\max}$ is maximized over the local polarization measurement directions.

For the $q\bar{q}\to gg$ channel, the concurrence and maximal CHSH parameter are
\begin{align}
    \mathcal C_{q\bar{q}\to gg}=-c_q(\theta),\quad
    S_{\max}^{q\bar{q}\to gg}=2\sqrt{1+c_q^2(\theta)}.
    \label{eq:entanglement-qqbar}
\end{align}
For $0<\theta<\pi$, $\mathcal C_{q\bar{q}\to gg}>0$ and $S_{\max}^{q\bar{q}\to gg}>2$; therefore, the final-state gluons are entangled throughout this angular range and can violate the CHSH inequality for optimal local polarization measurements.

\begin{figure}[htb]
\centering
\includegraphics[width=0.45\textwidth]{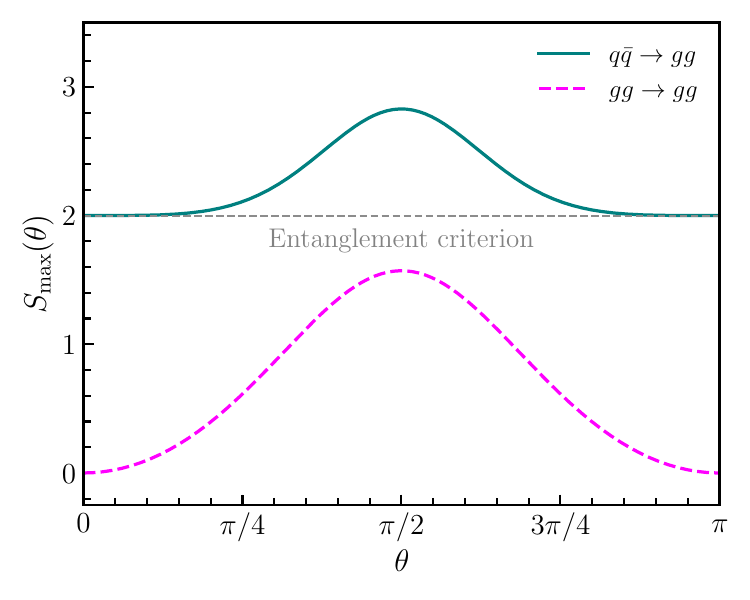}
\caption{The entanglement criterion $S_{\max}$ as a function of the scattering angle in different channels. The dashed line at $S_{\max} = 2$ separates the entangled states and separable states.}
\label{fig:entanglement_criterion}
\end{figure}

For the $gg\to gg$ channel, the concurrence and maximal CHSH parameter are
\begin{align}
    \mathcal C_{gg\to gg}=0,\quad
    S_{\max}^{gg\to gg}(\theta)
    =2\sqrt{c_z^2(\theta)+c_g^2(\theta)}.
    \label{eq:quantum-correlations-gg}
\end{align}
For all scattering angles, $\mathcal C_{gg\to gg}=0$ and $S_{\max}^{gg\to gg}\leq 10\sqrt{2}/9<2$, with the maximum attained at $\theta=\pi/2$. Therefore, the final-state gluons are separable throughout the angular range and do not violate the CHSH inequality for any choice of local polarization measurements. We show the numerical results of $S_{\max}$ in Fig.~\ref{fig:entanglement_criterion}.

However, in the polarized $pp$ collisions, the incoming gluons can be circularly polarized. In this case, we can control the helicity combination of incoming gluons. For instance, by reducing the $|++\rangle$ and $|--\rangle$ configurations for colliding gluons, we can significantly reduces $s$-channel contribution and enhance the entanglement of the final state gluon pair. More details are presented in Appendix \ref{sec:app-pol}. 

\section{E4C induced by linear polarization correlation of back-to-back gluon pair}
\label{sec:phenomenology}

\begin{figure}[htb]
    \centering
    \includegraphics[width=0.9\linewidth]{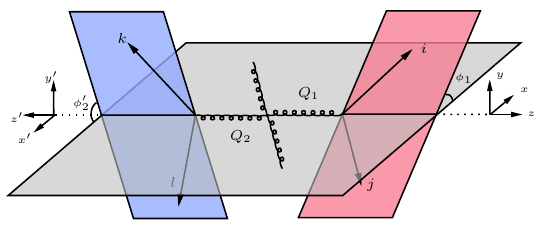}
    \caption{Illustration of $pp$ collisions using the channel $gg \to gg(\to q_i\bar{q}_jq_k\bar{q}_l)$ as an example. The pairs $(i,j)$ and $(k,l)$ belong to gluon initiated jets of energies $Q_1$ and $Q_2$, respectively, and define the corresponding branching planes. The azimuthal angles of the two branching planes with respect to the scattering plane are denoted by $\phi_1$ and $\phi_2^\prime$ defined in their local frame. }
    \label{fig:gg-to-qqbar}
\end{figure}

To connect the linear polarization correlation derived in Sec.~\ref{sec:partonic-density} with an experimental observable, we consider the azimuthal modulation of the four-point energy correlator. Following Ref.~\cite{Wang:2026uej}, the E4C correlates a particle pair in each of two gluon initiated jets while preserving information about the relative azimuthal orientation of the corresponding branching planes. It is defined as
\begin{align}
    \mathrm{E4C}(\Delta\Phi,n)
    =
    \sum_{ij,kl}\int\dfrac{\mathrm{d}\sigma}{\sigma}
    \dfrac{E_i^n E_j^n}{Q_1^{2n}}
    \dfrac{E_k^n E_l^n}{Q_2^{2n}}
    \delta\!\left(\Delta\Phi-(\phi_1-\phi_2^\prime)\right).
    \label{eq:e4c}
\end{align}
Here, the sums run over particle pairs $(i,j)$ in the first jet and $(k,l)$ in the second jet. The quantities $E_i$ and $E_j$ denote the energies of the particles in the first jet, whose energy is $Q_1$, while $E_k$ and $E_l$ denote the corresponding particle energies in the second jet, whose energy is $Q_2$. The normalized weights $E_i^nE_j^n/Q_1^{2n}$ and $E_k^nE_l^n/Q_2^{2n}$ are therefore invariant under independent rescalings of all energies in the two jets. The azimuthal asymmetry angle $\Delta\Phi=\phi_1-\phi'_2$, where $\phi_1$ is the azimuthal angle of the $(i,j)$-branching plane in the $xOy$ frame, while $\phi'_2$ is that of the $(k,l)$-branching plane in the $x'Oy'$ frame, as illustrated in Fig.~\ref{fig:gg-to-qqbar}. The exponent $n$ controls the energy weighting of the correlator. Notice that the azimuthal angles are always defined in the rest frame of the gluon pair instead of the lab frame, similar to that in Ref.~\cite{Bacchetta:2004it} in the language of dihadron fragmentation function.

In this work, we apply the E4C observable studied in Ref.~\cite{Wang:2026uej} to hard QCD $2\to2$ scattering. Contracting the spin dependent matrices of the two collinear splittings with the hard two gluon polarization density matrix maps the linear polarization correlation of the gluon pair onto a measurable azimuthal modulation in unpolarized proton-proton collisions. Whereas in $H\to gg$ the correlation reflects the structure of the color singlet $Hgg$ vertex, here it is determined by the helicity structure of the QCD scattering amplitudes. This feature is especially relevant in the $gg\to gg$ channel. Although each outgoing gluon is individually unpolarized, the pair retains a nonzero linear polarization correlation that cannot be accessed from either gluon alone but is probed by the E4C through the relative azimuthal angle of the two branching planes.

In the one-step collinear splitting approximation used below, the relevant partonic configuration factorizes into a hard subprocess, $gg\to g_1g_2$ or $q\bar{q}\to g_1g_2$, followed by the collinear splittings $g_1\to a_1a_2$ and $g_2\to b_1b_2$. Each daughter pair defines a splitting mode $\kappa_r\in\{q_f\bar{q}_f,gg\}$, where $r=1,2$ labels the two jets and $f$ denotes the quark flavor. Thus, $\boldsymbol{\kappa}=(\kappa_1,\kappa_2)$ labels the four parton configurations $q\bar{q} q\bar{q}$, $q\bar{q} gg$, including its jet interchanged configuration, and $gggg$. For a fixed splitting mode $\boldsymbol{\kappa}$, we denote the corresponding azimuthal distribution by $\mathrm d\Sigma_{\boldsymbol{\kappa}}/\mathrm d\Delta\Phi$ and define its normalization as $\Sigma_{\boldsymbol{\kappa}}=\int \mathrm d\Delta\Phi\,\mathrm d\Sigma_{\boldsymbol{\kappa}}/\mathrm d\Delta\Phi$.

The normalized differential distribution can then be written in the density matrix form
\begin{align}
    &\frac{\pi}{\Sigma_{\boldsymbol{\kappa}}}
    \frac{\mathrm d\Sigma_{\boldsymbol{\kappa}}}
    {\mathrm d\Delta\Phi\,\mathrm dz_1\,\mathrm dz_2}
    \nonumber\\
    &\quad =
    \frac{w_n(z_1,z_2)}{\mathcal N_{\boldsymbol{\kappa}}}
    \int_0^\pi\frac{\mathrm d\phi_1}{\pi}
    \sum_{\lambda_3,\lambda_4,\lambda_3',\lambda_4'}
    \rho^{\lambda_3\lambda_4,
    \lambda_3'\lambda_4'}
    \nonumber\\
    &\qquad\quad\times
    D_{g_1\to a_1a_2}^{\lambda_3'\lambda_3}
    (z_1,\phi_1)\,
    D_{g_2\to b_1b_2}^{\lambda_4'\lambda_4}
    (z_2,\phi_1 - \Delta\Phi).
    \label{eq:e4c-density-matrix-master}
\end{align}
Here, $\phi_1$ denotes the angle between the branching plane in the first jet and the scattering plane. The corresponding angle for the branching plane in the second jet is $\phi_1-\Delta\Phi$ where $\Delta\Phi$ is the relative azimuthal angle defined above. The normalization factor for a fixed splitting mode $\boldsymbol{\kappa}$ is
\begin{align}
    \mathcal N_{\boldsymbol{\kappa}}
    \equiv{}&
    \int_0^1\mathrm dz_1\,\int_0^1\mathrm dz_2\,
    w_n(z_1,z_2)
    P_{\kappa_1}(z_1)P_{\kappa_2}(z_2),
    \label{eq:e4c-topology-normalization}
\end{align}
where the E4C energy weight reads
\begin{align}
    w_n(z_1,z_2)\equiv[z_1(1-z_1)z_2(1-z_2)]^n. 
\end{align}
As in previous work, the gluon splitting matrix entering Eq.~\eqref{eq:e4c-density-matrix-master} is
\begin{align}
    D^{\lambda\lambda'}_{g\to ab}(z,k_\perp)
    &=
    \epsilon_{\lambda,\mu}\epsilon^*_{\lambda',\nu}
    P^{\mu\nu}_{ab\leftarrow g}(z,k_\perp),
    \\
    P^{\mu\nu}_{q\bar{q}\leftarrow g}(z,k_\perp)
    &=
    T_F\bigg[-g^{\mu\nu}
    +4z(1-z)\frac{k_\perp^\mu k_\perp^\nu}{k_\perp^2}\bigg],
    \\
    P^{\mu\nu}_{gg\leftarrow g}(z,k_\perp)
    &= 2N_c\bigg[ -g^{\mu\nu}\left(\frac{z}{1-z}+\frac{1-z}{z}\right)
    \nonumber\\
    &\qquad\qquad-2z(1-z)\frac{k_\perp^\mu k_\perp^\nu}{k_\perp^2} \bigg].
    \label{eq:tensor-splitting-kernels}
\end{align}
These tensors are the standard collinear splitting kernels~\cite{Gribov:1972ri,Dokshitzer:1977sg,Altarelli:1977zs}, with $T_F=1/2$. Here, $z$ is the longitudinal momentum fraction, and $k_\perp$ is the relative transverse momentum of the two daughter partons. After contraction with the gluon polarization vectors, the azimuthal dependence of the splitting matrix can be decomposed into an azimuthally independent term and a $\cos2 \phi$ anisotropic term. We denote the corresponding kernels by $P_{\kappa_r}(z)$ and $P_{\kappa_r}^{2\phi}(z)$, respectively. 

Integrating Eq.~\eqref{eq:e4c-density-matrix-master} over $z_1$ and $z_2$ and using the normalization in Eq.~\eqref{eq:e4c-topology-normalization}, we obtain
\begin{align}
    \frac{\pi}{\Sigma_{\boldsymbol{\kappa}}}
    \frac{\mathrm d\Sigma_{\boldsymbol{\kappa}}}
    {\mathrm d\Delta\Phi}
    =
    1+\mathscr{A}^{\boldsymbol{\kappa}}\cos(2\Delta\Phi).
    \label{eq:e4c-normalized-observable}
\end{align}
At leading power, the $\cos 2\Delta\Phi$ asymmetry $\mathscr{A}^{\boldsymbol{\kappa}}$ is obtained by combining the gluon pair polarization correlation with the spin dependent collinear splitting of each outgoing gluon. The resulting asymmetry is
\begin{align}
    \mathscr{A}^{\boldsymbol{\kappa}}(n,\theta)
    ={}&
    \frac{C_{xx'}(\theta)+C_{yy'}(\theta)}{2}\,a^{\boldsymbol{\kappa}}(n),
    \label{eq:resulting-asymmetry}
\end{align}
where $a^{\boldsymbol{\kappa}}(n)$ denotes the polarization analyzing power associated with the collinear splittings of the two outgoing gluons, determined by the spin dependent and unpolarized splitting kernels
\begin{align}
    a^{\boldsymbol{\kappa}}(n)
    =
    \frac{
        \displaystyle
        \int_0^1\mathrm dz_1\,\mathrm dz_2\,
        w_n(z_1,z_2)
        P_{\kappa_1}^{2\phi}(z_1)
        P_{\kappa_2}^{2\phi}(z_2)
    }{
        \displaystyle
        \int_0^1\mathrm dz_1\,\mathrm dz_2\,
        w_n(z_1,z_2)
        P_{\kappa_1}(z_1)
        P_{\kappa_2}(z_2)
    }.
    \label{eq:e4c-splitting-analyzer}
\end{align}

For inclusive four parton final states, the E4C azimuthal modulation receives contributions from the combinations $q\bar{q}q\bar{q},\ q\bar{q}gg,\ gggg$. As can be inferred from the results in Table~\Rmnum{1} of Ref.~\cite{Wang:2026uej}, cancellations among these contributions strongly suppress the inclusive $\cos 2\Delta\Phi$ signal and therefore dilute the sensitivity to the underlying gluon polarization correlation. To circumvent these cancellations, one may employ heavy flavor tagging, \textit{e.g.}, by requiring a tagged heavy quark pair in each of the two jets, $Q\bar{Q}$, with $Q=c,b$, thereby selecting the $Q\bar{Q}Q\bar{Q}$ channel. Such a selection removes the cancellations between quark and gluon splitting channels and can enhance the magnitude of the E4C observable by orders of magnitude, restoring its power as a clean probe of the gluon pair polarization correlation.

For the separable $gg\to gg$ state discussed in Sec.~\ref{subsec:Entanglement}, a nonzero E4C modulation should be interpreted as evidence of linear polarization correlation of the gluon pair rather than as evidence for entanglement or Bell nonlocality. At central scattering, $c_q(\pi/2)=C_{xx'}^{(q\bar{q})}(\pi/2)=-1$ and $c_g(\pi/2)=C_{xx'}^{(gg)}(\pi/2)=1/9$. Combining this value with $a^{(q\bar{q},q\bar{q})}(n=4)\simeq0.6944$ gives $\mathscr{A}_{q\bar{q}\to gg}^{(q\bar{q},q\bar{q})}(4,\pi/2)\simeq -0.6944$ and $\mathscr{A}_{gg\to gg}^{(q\bar{q},q\bar{q})}(4,\pi/2)\simeq7.7\times10^{-2}$.

Within the one-step collinear splitting approximation, requiring a tagged heavy quark pair in each jet selects events with two outgoing gluons. At Born level, the relevant subprocesses are $gg\to gg$ and $q\bar{q}\to gg$, whereas $qg\to qg$ and purely quark final states do not satisfy the flavor tag requirement. The \NNLOJET calculation of Ref.~\cite{Chen:2022tpk} shows that inclusive low-$p_T$, central dijet production is dominated by the $gg$ and $qg$ initial state channels, with a comparatively small $q\bar{q}$ contribution. Although that calculation does not impose the heavy flavor tagging requirement, the resulting channel hierarchy suggests that the tagged sample should be dominated by $gg\to gg$, with $q\bar{q}\to gg$ as the leading irreducible Born level contamination.

\section{Summary}
\label{sec:summary}

We study the polarization correlations of gluon pairs produced in the $q\bar{q}\to gg$ and $gg\to gg$ channels by constructing their complete polarization density matrices. Although the reduced polarization state of each outgoing gluon is unpolarized, the two gluon system retains nontrivial polarization correlations. The $q\bar{q}\to gg$ channel produces an entangled state which becomes maximally entangled at central scattering. By contrast, the $gg\to gg$ state remains separable and Bell local across the whole kinematic range, while still exhibiting a nonzero polarization correlation. We further show that the E4C probes these correlations through a characteristic $\cos 2\Delta\Phi$ modulation characterizing the relative azimuthal orientation of the two branching planes with respect to the scattering plane. Heavy flavor tagging can suppress cancellations among splitting channels and improve sensitivity to the correlation signal. 

\begin{acknowledgments}
    We thank Jian Zhou and Shuo Lin for helpful discussions. This work is supported by the National Natural Science Foundation of China under Grant No. 12405156, and the Shandong Province Natural Science Foundation under grant No.~2023HWYQ-011.
\end{acknowledgments}

\appendix

\section{Density matrix of the back-to-back gluon pair for longitudinally polarized initial states}
\label{sec:app-pol}

It is interesting to investigate the entanglement in polarized $pp$ collisions as well. For simplicity, we only study the case where the incoming partons are longitudinally polarized. The spin density matrix of the incoming parton $a$ can be written as
\begin{align}
    \rho_a^{\mathrm{in}}
    =
    \frac{1}{2}
    \left(
    \mathbb{I}+P_a\sigma_z
    \right),  
    \qquad a=1,2.
\end{align}
Here, $|P_a|\leq 1$ is the longitudinal polarization with $a$ labeling the parton flavor.

Following Ref.~\cite{Pankov:2002qk}, we define a parameter $\omega_{\lambda_1\lambda_2}$ to quantify the probability of the initial helicity configuration $|\lambda_1\lambda_2\rangle$. Assuming that the two colliding partons are uncorrelated, it is given by
\begin{align}
    \omega_{\lambda_1\lambda_2}
    =\frac{(1+\lambda_1P_1)(1+\lambda_2P_2)}{4},
    \quad \lambda_1,\lambda_2=\pm1.
\end{align}
These weights satisfy $\omega_{\lambda_1\lambda_2}\geq0$ and $\sum_{\lambda_1,\lambda_2}\omega_{\lambda_1\lambda_2}=1$. The corresponding unnormalized final-state hard matrix thus reads
\begin{align}
    H_{\alpha \beta}(P_1,P_2) = \sum_{\lambda_1,\lambda_2}\omega_{\lambda_1\lambda_2}
    \mathcal{M}_{\lambda_1\lambda_2;\alpha}\mathcal{M}_{\lambda_1\lambda_2;\beta}^*
\end{align}
Here, $\alpha,\beta=1,\ldots,4$ label the ordered final-state two gluon helicity basis $\{|++\rangle,\ |+-\rangle,\ |-+\rangle,\ |--\rangle\}$. Thus, $\mathcal M_{\lambda_1\lambda_2;\alpha}$ denotes the amplitude for a fixed initial helicity configuration $(\lambda_1,\lambda_2)$ to produce the $\alpha$-th final-state helicity state. \footnote{We note a typographical error in Eq.~(10.139) of Ref.~\cite{Gastmans:1990xh}, where the variables $t$ and $u$ should be interchanged. The expressions in Eqs.~(5.21) and (5.22) of the same reference are correct, as the definitions of $t$ and $u$ are reversed in that section.}

For the $q\bar{q} \to gg$ channel, the normalized density matrix is
\begin{align}
    \rho_{q\bar q\to gg}
    =
    \begin{pmatrix}
        0 & 0 & 0 & 0\\
        0 & \rho_{22}^{q\bar q}
          & \rho_{23}^{q\bar q} & 0\\
        0 & \rho_{32}^{q\bar q}
          & \rho_{33}^{q\bar q} & 0\\
        0 & 0 & 0 & 0
    \end{pmatrix}.
\end{align}
The nonzero matrix elements are
\begin{align}
    \rho_{22}^{q\bar q}
    &=
    \frac{1}{2}
    +
    \frac{
    (P_q-P_{\bar q})\cos\theta}{
    (1-P_qP_{\bar q})
    (1+\cos^2\theta)},
    \\
    \rho_{33}^{q\bar q}
    &=
    \frac{1}{2}
    -
    \frac{
    (P_q-P_{\bar q})\cos\theta}{
    (1-P_qP_{\bar q})
    (1+\cos^2\theta)},
    \\
    \rho_{23}^{q\bar q}
    &=
    \rho_{32}^{q\bar q}
    =
    -\frac{
    \sin^2\theta}{
    2(1+\cos^2\theta)}.
\end{align}
Although the diagonal elements depend on $P_q$ and $P_{\bar q}$, the normalized off-diagonal ones are polarization independent. Consequently, the normalized two-gluon correlation tensor, concurrence, and maximal CHSH parameter remain identical to those for the unpolarized initial state. The initial-state polarization shifts the diagonal populations of the $|+-\rangle$ and $|-+\rangle$ states in opposite directions, thereby in general inducing equal and opposite longitudinal polarizations of the two final-state gluons,
\begin{align}
    B_z^{g_3}=-B_{z'}^{g_4}= 
    \frac{2\cos\theta\,(P_q-P_{\bar q})}
    {(1+\cos^2\theta)(1-P_qP_{\bar q})}.
\end{align}
Here, we have chosen $\theta$ as the angle between $g_3$ and quark.

For the $gg\to gg$ channel, the normalized two gluon polarization density matrix
can be written as
\begin{align}
    \rho_{gg\to gg}
    =
    \frac{1}{N_{gg}}
    \begin{pmatrix}
        h_{11} & 0 & 0 & 0\\
        0 & h_{22} & h_{23} & 0\\
        0 & h_{32} & h_{33} & 0\\
        0 & 0 & 0 & h_{44}
    \end{pmatrix},
\end{align}
where the nonzero elements of the unnormalized hard matrix are
\begin{align}
    h_{11}
    &=
    (1+P_{g_1})(1+P_{g_2}),
    \\
    h_{44}
    &=
    (1-P_{g_1})(1-P_{g_2}),
    \\
    h_{22}
    &=
    \frac{1}{8}
    \Big[
    (1-P_{g_1}P_{g_2})
    (1+6\cos^2\theta+\cos^4\theta)
    \nonumber\\
    &\qquad\;
    +4(P_{g_1}-P_{g_2})
    \cos\theta(1+\cos^2\theta)
    \Big],
    \\
    h_{33}
    &=
    \frac{1}{8}
    \Big[
    (1-P_{g_1}P_{g_2})
    (1+6\cos^2\theta+\cos^4\theta)
    \nonumber\\
    &\qquad\;
    -4(P_{g_1}-P_{g_2})
    \cos\theta(1+\cos^2\theta)
    \Big],
    \\
    h_{23}
    &=
    h_{32}
    =
    \frac{
    (1-P_{g_1}P_{g_2})\sin^4\theta
    }{8}.
\end{align}
The normalization factor is 
\begin{align}
    N_{gg}
    &=2\bigl(1+P_{g_1}P_{g_2}\bigr)
    \nonumber
    \\
    &
    +\frac14\bigl(1-P_{g_1}P_{g_2}\bigr)
    \bigl(1+6\cos^2\theta
    +\cos^4\theta\bigr).
\end{align} 

The polarized initial state can also induce nonvanishing longitudinal polarizations of the individual final-state gluons,
\begin{align}
    B_z^{g_3}
    &=
    \frac{
    2(P_{g_1}+P_{g_2})
    +\cos\theta(1+\cos^2\theta)(P_{g_1}-P_{g_2})
    }{N_{gg}},
    \\
    B_{z'}^{g_4}
    &=
    \frac{
    2(P_{g_1}+P_{g_2})
    -\cos\theta(1+\cos^2\theta)(P_{g_1}-P_{g_2})
    }{N_{gg}}.
\end{align}

By contrast, for $gg\to gg$, both the correlation tensor and the entanglement properties of the final state depend explicitly on $P_{g_1}$ and $P_{g_2}$. The transverse linear polarization components of the two gluon correlation tensor are
\begin{align}
    C_{xx'}=C_{yy'} = 
    \dfrac{(1-P_{g_1}P_{g_2})
    \sin^4\theta}{4N_{gg}}. 
\end{align}
Unlike the unpolarized case, for which the final-state gluon pair is always separable, sufficiently strong longitudinal polarization of the initial gluons can generate an entangled final state.

As an illustrative case, consider central scattering $\theta=\pi/2$ with equal and opposite initial polarizations, $P_{g_2}=-P_{g_1}$. In this configuration, the longitudinal polarizations of the individual final-state gluons vanish, $B_z^{g_3}=B_{z'}^{g_4}=0$, whereas the transverse correlations become
\begin{align}
    C_{xx'}=C_{yy'}
    =
    \frac{1+P_{g_1}^{2}}
    {9-7P_{g_1}^{2}}.
\end{align}
They increase monotonically from \(1/9\) in the unpolarized case to unity in the fully polarized limit.

The concurrence correspondingly becomes
\begin{align}
    \mathcal C_{gg\to gg}
    =
    \max\left[0,
    \frac{
    9P_{g_1}^{2}-7
    }{
    9-7P_{g_1}^{2}
    }
    \right].
    \label{eq:concurrence-polarized-gg-central}
\end{align}
Hence, the final-state gluon pair becomes entangled when
\begin{align}
    |P_{g_1}|=|P_{g_2}|
    >
    \frac{\sqrt{7}}{3}.
\end{align}
For fully and oppositely polarized initial gluons, $|P_{g_1}|=|P_{g_2}|=1$, the concurrence reaches unity and the final-state density matrix reduces to $\rho_{gg\to gg} = |\Psi^+\rangle\langle\Psi^+|$, corresponding to a maximally entangled Bell state.

\end{document}